\documentclass[11pt]{article}

\usepackage[final]{acl}

\usepackage{times}
\usepackage{latexsym}
\usepackage{booktabs}
\usepackage{array}
\usepackage{xltabular}   
\usepackage{float}
\usepackage{amsmath}
\usepackage{hyperref}
\usepackage{tcolorbox}
\hypersetup{colorlinks=true, linkcolor=blue!60!black, citecolor=blue!60!black,
            urlcolor=blue!60!black}
\usepackage[capitalize, noabbrev]{cleveref}
\usepackage{cuted}
\usepackage{url}

\usepackage[T1]{fontenc}

\usepackage[utf8]{inputenc}

\usepackage{microtype}

\usepackage{inconsolata}

\usepackage{graphicx}
\usepackage{seqsplit}

\usepackage{listings}
\usepackage{xcolor}
\usepackage{subcaption}

\definecolor{codeblue}{RGB}{30,80,160}
\definecolor{codered}{RGB}{165,55,55}
\definecolor{codebrown}{RGB}{150,90,25}
\definecolor{codepink}{RGB}{150,80,110}
\definecolor{codeframe}{RGB}{190,190,190}
\definecolor{codelineno}{RGB}{150,150,150}
\definecolor{codebg}{RGB}{250,250,250}

\lstdefinelanguage{yamlkb}{
  sensitive=true,
  morecomment=[l]{\#},
  morestring=[b]",
  morestring=[b]',
  alsoletter={_},
  keywords={true,false,null,True,False,None,yes,no},
  keywordstyle=\color{codeblue}\bfseries,
}

\lstdefinestyle{kbstyle}{
  language=yamlkb,
  basicstyle=\ttfamily\footnotesize,
  commentstyle=\color{codepink},
  stringstyle=\color{codered},
  identifierstyle=\color{black},
  numbers=left,
  numberstyle=\ttfamily\scriptsize\color{codelineno},
  numbersep=8pt,
  stepnumber=1,
  frame=single,
  rulecolor=\color{codeframe},
  framerule=0.4pt,
  framesep=4pt,
  xleftmargin=18pt,
  framexleftmargin=18pt,
  backgroundcolor=\color{codebg},
  showstringspaces=false,
  columns=fullflexible,
  keepspaces=true,
  breaklines=true,
  breakatwhitespace=false,
  postbreak=\mbox{\textcolor{codeblue}{$\hookrightarrow$}\space},
  emph={[1]benchmark,case_defaults,case_grid,metrics,plots,
        warmup,repeat,use_cuda_graph,flush_l2,autotune,
        M,N,dtype,flops_expr,bytes_expr,
        latency_ms,bandwidth_GBs,speedup,fp16,fp32},
  emphstyle={[1]\color{codeblue}\bfseries},
}
\usepackage{tcolorbox}
\tcbuselibrary{listings,skins,breakable}

\lstdefinelanguage{plainprompt}{
  keywords={},
  sensitive=true
}

\lstdefinestyle{promptstyle}{
  language=plainprompt,
  basicstyle=\ttfamily\footnotesize\color{black},
  identifierstyle=\color{black},
  keywordstyle=\color{black}\mdseries,
  commentstyle=\color{black}\mdseries,
  stringstyle=\color{black}\mdseries,
  emphstyle={[1]\color{black}\mdseries},
  literate={0}{{0}}{1}{1}{{1}}{1}{2}{{2}}{1}{3}{{3}}{1}{4}{{4}}{1}{5}{{5}}{1}{6}{{6}}{1}{7}{{7}}{1}{8}{{8}}{1}{9}{{9}}{1},
  numbers=left,
  numberstyle=\ttfamily\scriptsize\color{gray},
  numbersep=8pt,
  stepnumber=1,
  frame=none,
  columns=fullflexible,
  keepspaces=true,
  breaklines=true,
  breakatwhitespace=false,
  showstringspaces=false
}

\newtcblisting{promptbox}[1][]{%
  enhanced,
  breakable,
  colback=black!1,
  colframe=black!35,
  coltitle=black,
  fonttitle=\bfseries\small,
  colbacktitle=black!7,
  boxrule=0.5pt,
  arc=1.5pt,
  left=6pt,
  right=6pt,
  top=4pt,
  bottom=4pt,
  before skip=6pt,
  after skip=6pt,
  listing only,
  listing options={style=promptstyle},
  title={#1}
}

\title{TileBench: A Controlled Benchmark for Performance Evaluation and Bottleneck Diagnosis of Tile-Based Programming Models}

\author{
 \textbf{Bowen Cui\textsuperscript{1,*}},
 \textbf{Zhongchun Zhou\textsuperscript{1,*}},
 \textbf{Hao Wu\textsuperscript{1}},
 \textbf{Tejas Ramesh\textsuperscript{1}},
\\
 \textbf{Junyu Yin\textsuperscript{1}},
 \textbf{Jialiang Gu\textsuperscript{1}},
 \textbf{ Keren Zhou\textsuperscript{1}}
\\
\\
 \textsuperscript{1}George Mason University
\\
 \small{
   \textbf{Correspondence:} \href{mailto:kzhou6@gmu.edu}{kzhou6@gmu.edu}
 }
}

\begin{document}
\maketitle
\renewcommand{\thefootnote}{\fnsymbol{footnote}}
\footnotetext[1]{These authors contributed equally to this work.}
\renewcommand{\thefootnote}{\arabic{footnote}}
\begin{abstract}
Tile-based programming models, such as Triton and cuTile, aim to simplify high-performance kernel development, but their practical performance, tuning behavior, and usability remain difficult to compare systematically.
We present TileBench, a controlled benchmark for evaluating Triton and cuTile on NVIDIA B200 GPUs under matched operator semantics and comparable implementation structures.
TileBench contains 45 operators covering diverse AI-kernel patterns and memory/computation behaviors.
Each task provides a PyTorch reference, verified Triton and cuTile implementations, standardized data-types (dtype) and input-size sweeps, default and autotuned configurations, roofline-based metrics, and profiling-guided diagnosis.
Our evaluation shows that performance gaps are workload-dependent: cuTile excels on a small cluster of Tensor-Core/TMA-friendly kernels, while Triton is stronger on many irregular, streaming, and bandwidth-bound operators. 
We further evaluate LLM-generated cuTile and Triton kernels and find that Triton is consistently more token-efficient than cuTile under the same iterative refinement protocol.
TileBench is publicly available at \url{https://github.com/Deep-Learning-Profiling-Tools/Tilebench}.

\end{abstract}

\section{Introduction}

Modern AI workloads, such as Large Language Models (LLMs)~\cite{llm}, diffusion models~\cite{diffusion_model}, and multi-modal models~\cite{multi-model}, increasingly depend on custom kernels beyond standard library primitives, including attention variants~\cite{flashattention2}, normalization~\cite{rmsnorm}, Mixture-of-Experts (MOE) routing~\cite{moe}, quantization and dequantization~\cite{quantization}, and fused point-wise operations~\cite{sol-excebench}, which involve diverse computation patterns.
Although highly optimized implementations for common primitives are available in vendor libraries such as cuBLAS~\cite{cublas} and cuDNN~\cite{cudnn}, template libraries such as CUTLASS~\cite{cutlass}, and specialized kernels such as FlashAttention~\cite{flashattention2}, many emerging operators still require custom kernel development involving custom fusion patterns, memory layouts, and advanced algorithms to fully exploit modern GPU hardware~\cite{cutile,sol-excebench}.

Implementing high-performance GPU kernels remains difficult.
CUDA~\cite{cuda} provides fine-grained control over the memory hierarchy, synchronization, and Tensor Cores, but this control comes with substantial engineering costs~\cite{kernelevolve}.
High-level tile-based programming models such as cuTile~\cite{cutile} and Triton~\cite{triton} aim to reduce this burden while offering efficient performance.
Unlike the traditional Single Instruction Multiple Threads (SIMT) programming model~\cite{SIMT}, these domain-specific languages (DSLs) enable developers to program with a higher abstraction: they load and store chunks of data (Tiles) and apply certain computations on these tiles through tile-level primitives, facilitating the adoption of Tensor Cores (i.e., \texttt{mma}), Tensor Memory Accelerator (i.e., \texttt{cp.async.bulk.tensor}), and Tensor Memory instructions (i.e., \texttt{tcgen05}). 
The design philosophy of these DSLs enables developers to "focus on your algorithm" while leaving complex code optimizations to the compiler and runtime. 




Though such abstractions hide low-level complexity, it is still difficult for developers to diagnose performance bottlenecks and optimize the kernels for diverse AI operators.
Recent cuTile evaluation studies~\cite{yadav2026evaluating} compare cuTile with cuBLAS, Triton~\cite{triton}, Warp Matrix Multiply and Accumulate (WMMA), and raw SIMT on general matrix multiplication (GEMM) and attention across Hopper and Blackwell GPUs, showing that cuTile can be highly effective on selected B200 workloads but is also strongly workload- and architecture-dependent. 
On the other hand, Triton can yield different performance with regard to different versions of the compiler, autotuning space, Machine Learning (ML) algorithm, or implementation details when integrated into systems and libraries such as vLLM~\cite{vllm}, LightLLM~\cite{lightllm}, Liger-Kernel~\cite{liger-kernel}, and Unsloth~\cite{unsloth}.
As a result, the community lacks a systematic understanding of performance bottlenecks, best practices for coding with these DSLs, and comparisons among different tile-centric DSLs.

To support fair evaluation and future tooling for tile-centric DSLs, we present \textbf{TileBench}, a controlled benchmark for comparing Triton and cuTile implementations of AI operators.
TileBench contains 45 operators spanning attention and sequence primitives, GEMM variants, normalization, reductions, loss functions, elementwise and fused pointwise kernels, convolution and stencil computations, data movement, indexing, quantization, and sorting. 
Each task includes a PyTorch reference implementation for correctness checking and baseline measurement, together with matched Triton and cuTile implementations. 

Our work makes the following contributions:


\begin{itemize}
    \item We introduce TileBench, a controlled benchmark of 45 GPU operators on an NVIDIA B200 GPU, with manually written and verified Triton/cuTile implementations under matched semantics, comparable implementation structures, and standardized multi-dtype and input-size sweeps.
    

    \item We evaluate both default and autotuned configurations, quantifying how much performance is gained from systematic autotuning and how much performance headroom remains relative to hardware roofline limits.

    \item We build a profiling-guided analysis workflow to explain performance differences between Triton and cuTile using generated code, compiler behavior, and hardware-level metrics.

    \item We evaluate Triton and cuTile as target programming models for LLM-generated kernels, measuring not only correctness and runtime performance but also the token consumption required to achieve performant implementations.

\end{itemize}

Across the suite, both programming languages substantially outperform PyTorch, while Triton achieves stronger overall performance, and cuTile is competitive on a small set of Tensor Core and TMA-friendly workloads.
Autotuning improves both backends but leaves substantial roofline headroom, suggesting that compiler lowering and memory behavior often dominate tile-parameter choices.
Evaluation on LLM-generated cuTile and Triton kernels demonstrates that Triton is consistently more token-efficient than cuTile under the same iterative refinement protocol.


\section{Related Work}

\subsection{Tile-Based GPU Programming Models}


Triton~\cite{triton} and cuTile~\cite{cutile} are two representative tile-centric interfaces that are directly relevant to NVIDIA GPUs.
Triton exposes a Python-embedded block program in which users specify tile sizes, pointer arithmetic, and scheduling hints, while the compiler lowers tile-level tensor operations to GPU code.
On NVIDIA GPUs with Tensor Memory Accelerator (TMA) support, Triton's tensor descriptor API can lower descriptor-based loads and stores to TMA-backed memory~\cite{tritonMakeTensorDescriptor}.
cuTile raises a similar abstraction inside the CUDA ecosystem:
kernels are expressed through tile operations such as \texttt{ct.load}, \texttt{ct.mma}, and \texttt{ct.store}, and the compiler and runtime map these operations to threads, memory movement, Tensor Cores, and TMA.

The broader ecosystem also includes several production and research systems, such as TileLang~\cite{tilelang}, ThunderKittens~\cite{thunderkittens}, Tilus~\cite{tilus}, TLX~\cite{tlx}, and AWS NKI~\cite{nki}. 
However, these systems are not all directly comparable in our setting: NKI targets non-GPU accelerators, TLX extends Triton rather than serving as an independent high-level backend, and TileLang, ThunderKittens, and Tilus expose low-level parallelism, scheduling, layouts, and runtime controls.
TileBench, therefore, focuses on Triton and cuTile:
both can be evaluated on the same NVIDIA B200 platform, both expose tile-centric programming interfaces, and both support a meaningful apples-to-apples comparison under matched semantics, comparable tuning budgets, roofline analysis, and profiling-guided diagnosis.

\subsection{Benchmarks for GPU Kernel Generation and Optimization}

Recent benchmarks evaluate LLM-written or automatically optimized GPU kernels from different stages.
KernelBench~\cite{kernelbench} and CUDABench~\cite{cudabench} focus on text-to-CUDA generation;
BackendBench~\cite{backendbench} and TritonBench~\cite{tritonbench} study PyTorch backend or Triton operator generation;
FlashInfer-Bench~\cite{flashinferbench} emphasizes serving-oriented inference primitives;
ComputeEval~\cite{computeeval} covers broader CUDA programming ability;
and SOL-ExecBench~\cite{sol-excebench} introduces hardware speedup-of-light evaluation.
These benchmarks measure correctness, software-baseline speedup, GPU efficiency, or hardware headroom, but they do not isolate programming-model effects between tile-level DSLs.
TileBench is complementary; its main track uses manually written and verified Triton/cuTile implementations under matched operator semantics and comparable autotuning ranges, while \Cref{subsec:iteration} separately measures how easy and token-efficient it is for LLMs to generate correct and performant kernels for each DSL.

\section{TileBench}
\label{tilebench}

TileBench is a controlled benchmark for comparing Triton and cuTile under matched operator semantics and standardized evaluation. 
PyTorch serves as the semantic reference, correctness oracle, and software baseline, while Triton and cuTile are the primary programming models under comparison.

\subsection{Overview and Pipeline}
\label{overview_and_pipeline}

\begin{figure*}[t] 
  \centering
  \includegraphics[width=0.85\textwidth]{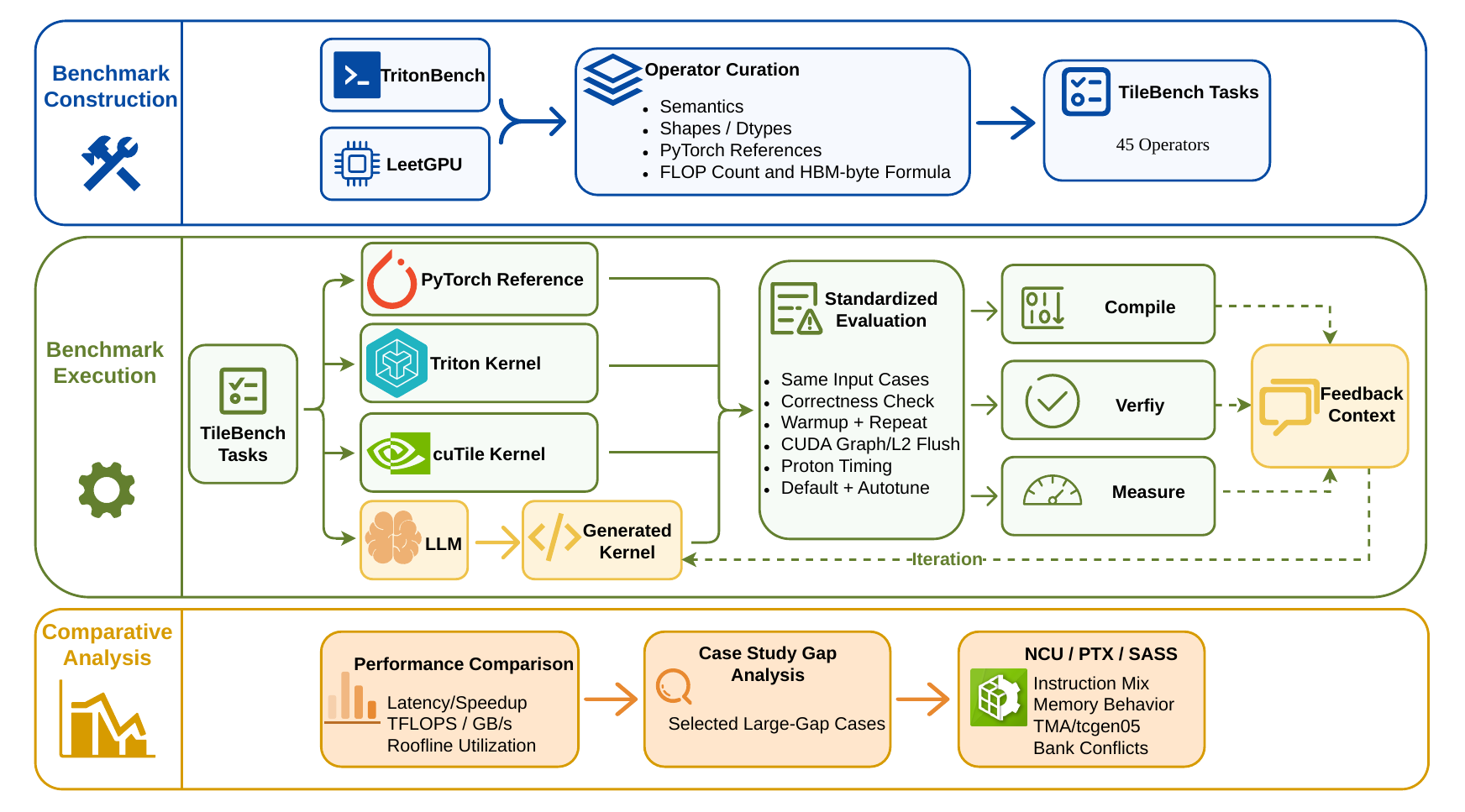}
  \caption{TileBench Overview.}
  \label{fig:tilebench_overview}
\end{figure*}

\cref{fig:tilebench_overview} shows the TileBench workflow.
The pipeline has three stages.
First, the benchmark construction stage curates operators from TritonBench~\cite{tritonbench} and LeetGPU~\cite{leetgpu} and normalizes them into a unified TileBench task format.
Second, the benchmark-execution stage evaluates each task through a shared harness:
PyTorch produces reference outputs and baseline latency, while Triton and cuTile implementations are compiled, run through the configured default or autotuned path, verified, and timed on the same input cases.
Third, the comparative analysis stage aggregates latency, speedup, throughput, effective bandwidth, and roofline utilization.
For selected large-gap cases, we further use Nsight Compute~\cite{ncu} together with PTX/SASS~\cite{ptx} inspection to analyze instruction mix, memory behavior, TMA/\texttt{tcgen05} usage, and bank conflicts.

\subsection{Benchmark Suite}
\label{benchmark_suite}

\paragraph{Benchmark Sources and Coverage.}
TileBench contains 45 representative AI operator tasks derived from existing benchmark suites, including 26 from \textsc{TritonBench}~\cite{tritonbench} and 19 from \textsc{LeetGPU}~\cite{leetgpu}.
We use these benchmarks as sources of operator semantics and task definitions, but all benchmark implementations are manually written and verified within the TileBench framework.
For each operator, we provide controlled benchmark artifacts including \texttt{impl\_torch.py}, \texttt{impl\_triton.py}, \texttt{impl\_cutile.py}, and \texttt{config.yaml}.
This design ensures matched code shapes, algorithms, uniform data types (dtype) and input shapes, shared timing protocols, and consistent autotuning configurations across implementations.
The main benchmark track evaluates manually implemented and verified Triton and cuTile kernels under the TileBench harness, while LLM-generated implementations are evaluated separately in \Cref{subsec:iteration}.

The operators are grouped into five categories:
\emph{Point-wise}, \emph{Reduction/Normalization}, \emph{Matrix Multiplication/Attention}, \emph{Stencil/Convolution}, and \emph{Data Layout}.
The category distribution is shown in \Cref{tab:tilebench_categories}.
The full per-operator list is reported in \Cref{appsubsec:full_benchmark_list}.

Dtype support and input-size sweeps are operator-specific. 
Each operator declares its supported dtypes and swept parameters in \texttt{config.yaml} (\Cref{appsubsec:config.yaml_example}), and TileBench evaluates 20 input configurations per supported dtype. 
This provides each task with standardized multi-dtype and multi-size coverage, designed to capture small-to-large scaling, B200 roofline behavior, and LLM-like regimes rather than inheriting the fixed cases from TritonBench or LeetGPU.



\paragraph{Standardization across Benchmarks.}
Standardization is necessary because the source benchmarks use different evaluation protocols: TritonBench provides Triton-focused operator-generation tasks and performance scripts, while LeetGPU follows a challenge-style format with per-problem functional tests and a single large performance case. 
TileBench reconstructs these operators into a unified benchmark with matched Triton/cuTile implementations, shared correctness checks, standardized multi-dtype and input-size sweeps, common timing controls, and comparable autotuning.

First, every operator has a PyTorch reference implementation that defines the semantic ground truth.
Triton and cuTile outputs are compared against this reference using a shared correctness checker:
floating-point outputs use \texttt{torch.testing.assert\_close} with dtype- and operator-specific tolerances, while integer outputs use exact comparison (\texttt{atol = rtol =0}).

Second, each operator has a \texttt{config.yaml} file (\Cref{appsubsec:config.yaml_example}) that declares the dtype list, case grid, verification tolerances, timing options, and analytical FLOP/HBM-byte formulas. 
The same formulas are evaluated for every \texttt{(case, dtype)} pair to derive throughput, effective bandwidth, arithmetic intensity, and roofline utilization consistently across backends.

Third, all implementations are measured by the same timing protocol and report the mean kernel latency.
Measurement uses the Proton profiler~\cite{proton} with CUDA graph capture-and-replay, L2-cache flushing, 20 warmup runs, and 100 timed runs; NVIDIA Nsight Compute (NCU) is reserved for selected post-hoc case studies.

Fourth, in the main benchmark track, the Triton and cuTile implementations provide both a default execution path and an autotuned execution path.
The default path uses a single fixed, manually selected configuration, while the autotuned path searches over a manually defined set of candidate configurations and records the winning one for analysis.
Since Triton and cuTile expose different tuning knobs, TileBench uses comparable tuning ranges rather than a one-to-one parameter mapping.
This design allows each backend to use its native tuning interface while keeping the autotuning budget and search granularity comparable across programming models.

\subsection{Iterative LLM Code-Generation}
\label{subsec:iteration}

TileBench further defines an iterative LLM code-generation track over the same operator set.
For each operator, the LLM generates Triton and cuTile implementations from the natural-language operator description and TileBench task specification;
it does not see or optimize the human-written implementations used in the main benchmark track.
Unlike the main benchmark track, LLM-generated code is evaluated without backend autotuning, and each iteration must commit to a single configuration.
Each backend is evaluated for up to $10$ refinement iterations on one representative large case per supported dtype, with verification and performance feedback used to guide the next implementation and configuration.
A backend is frozen once its implementation passes verification and reaches at least $80\%$ roofline utilization;
otherwise, refinement continues until the $10$-iteration budget is exhausted.
The formal definition of the roofline utilization used for this stopping criterion is provided in \Cref{appsubsec:evaluate_modes}.

At iteration 0, the prompt (\Cref{appsubsec:iter0_prompt}) is assembled from the TileBench framework guide (\Cref{appsubsec:framework_guide}), the Triton and cuTile API references (\Cref{appsubsec:api_reference}), the natural-language operator description (\Cref{appsubsec:problem_description}), the operator \texttt{config.yaml} (\Cref{appsubsec:config.yaml_example}), the PyTorch reference implementation (\Cref{appsubsec:torch_reference}), and the strict output-format instructions (\Cref{appsubsec:output_format}). 
The framework guide defines the implementation contract, multi-dtype handling, and hardware constraints.
For iteration $N>0$, the prompt (\Cref{appsubsec:iterN_prompt}) is rebuilt with the previous trajectory, per-backend configuration, verification status, roofline utilization, speedup over PyTorch, worst-first performance summary, previous generated source files, and, when the latest attempt regresses or fails verification, the best verify-clean source files as a rebuild reference.

The evaluator rejects three classes of reward hacking before timing:
delegating computation to PyTorch, cuBLAS, cuDNN, or the PyTorch reference implementation;
caching outputs using tensor identity, data pointer, or version-dependent keys;
and bypassing the intended search process by importing backend autotuners.
These cases are checked using static scans, two-pass anti-cache verification, and rotating-input timing.
Any detected failures are returned to the LLM as structured feedback in the next iteration.

Triton and cuTile are tracked independently during refinement.
The final result for each backend is selected as the best verify-clean iteration among all iterations executed before freezing or budget exhaustion.

\subsection{Evaluation Metrics}

\paragraph{Performance Metrics.}
For each operator \(o\), backend \(b\), dtype \(d\), and input case \(c\), the benchmark records mean kernel latency \(T_{o,b,d,c}\).
We report speedup over PyTorch as \(T_{o,\mathrm{torch},d,c}/T_{o,b,d,c}\), along with achieved throughput, effective bandwidth, arithmetic intensity, and roofline utilization.
Let \(F_{o,d,c}\) and \(B_{o,d,c}\) be the analytical FLOP count and HBM-byte count declared in \texttt{config.yaml}.
Achieved throughput is computed as
\[
\mathrm{TFLOPS}_{o,b,d,c}
=
\frac{F_{o,d,c}}{T_{o,b,d,c}}\cdot 10^{-12}.
\]
Arithmetic intensity is
$AI_{o,d,c}=F_{o,d,c}/B_{o,d,c}$.
Roofline utilization normalizes achieved throughput by the dtype-specific roofline bound:
\[
R_{o,b,d,c}
=
\frac{
\mathrm{TFLOPS}_{o,b,d,c}
}{
\min(P^{\mathrm{peak}}_{d}, AI_{o,d,c}\cdot BW^{\mathrm{peak}})
}.
\]
Here \(P^{\mathrm{peak}}_{d}\) denotes the B200 peak throughput for dtype \(d\), and \(BW^{\mathrm{peak}}\) denotes the B200 peak HBM bandwidth; both constants are specified in \texttt{B200.json} (\Cref{appsubsec:B200_JSON_Information}).

\paragraph{Token-aware Metrics.}
For the iterative LLM code-generation track, let $o$ denote an operator, $b\in\{\mathrm{Triton},\mathrm{cuTile}\}$ a backend, $i \in \{0,\ldots,9\}$ a refinement iteration, and $\mathcal{D}_o$ the supported dtypes evaluated in this track.
For each $(o,b,i)$, it records the backend-level token usage $C^{\mathrm{tok}}_{o,b,i}$, counting failed iterations because compilation failures and incorrect generations are part of the search process.

We use correctness-gated speedup to measure useful performance.
Let $\mathrm{correct}_{o,b,i}$ indicate that backend $b$ at iteration $i$ compiles and passes correctness checks for all evaluated dtypes.
For dtype $d$, let $T_{o,\mathrm{torch},d}$ be the PyTorch reference latency and $T_{o,b,i,d}$ be the generated implementation latency on the representative large case.
The correctness-gated speedup of iteration $i$ is
\[
G_{o,b,i}
=
\begin{cases}
\displaystyle
\frac{1}{|\mathcal{D}_o|}
\sum_{d \in \mathcal{D}_o}
\frac{T_{o,\mathrm{torch},d}}{T_{o,b,i,d}},
&
\text{if } \mathrm{correct}_{o,b,i},\\[8pt]
0,
&
\text{otherwise.}
\end{cases}
\]

Thus, incorrect or non-compiling iterations contribute zero speedup while still incurring token costs.
We report:
\[
\mathrm{BestSpeedup@10}_{o,b}
=
\max_{0 \le i < 10}
G_{o,b,i}.
\]
The token cost is accumulated over the $10-$iteration budget:
\[
\mathrm{TokenCost@10}_{o,b}
=
\sum_{i=0}^{9}
C^{\mathrm{tok}}_{o,b,i}.
\]

Token efficiency is the best correctness-gated speedup per million tokens:
\[
\begin{split}
&\mathrm{TokenEfficiency@10}_{o,b}
\\
&\qquad=
\frac{
\mathrm{BestSpeedup@10}_{o,b}
}{
\mathrm{TokenCost@10}_{o,b}/10^{6}
}.
\end{split}
\]
When reporting aggregate results, we macro-average $\mathrm{BestSpeedup@10}$, $\mathrm{TokenCost@10}$, and $\mathrm{TokenEfficiency@10}$ across operators for each backend.


\section{Experimental Setup}
\label{experimental_setup}

We run two evaluations on TileBench: a main benchmark track comparing the human-written Triton and cuTile implementations of all 45 operators, and the iterative LLM code-generation track described in \Cref{subsec:iteration}.

\paragraph{Hardware.}
All measurements are taken on a single NVIDIA B200 GPU with 180\,GB of HBM3e memory;
the measured peak bandwidth used for roofline analysis is $6539.4$\,GB/s.

\paragraph{Software and Sampling.}
We use PyTorch 2.10 with CUDA 13.0, Triton 3.6.0, and the \texttt{cuda-tile} 1.3.0 release.
Per-operator FLOP count and HBM-byte counts are evaluated from the \texttt{flops\_expr} and \texttt{bytes\_expr} declared in each \texttt{config.yaml}, ensuring throughput, bandwidth, and roofline utilization are derived consistently across backends.
For aggregated metrics, unless stated otherwise, we use geometric mean over cases within each operator and over operators.

\paragraph{Models.}
For the iterative LLM track we evaluate two reasoning models: OpenAI GPT-5.5~\cite{gpt5_5} and Anthropic Claude Opus 4.7~\cite{claudeopus4_7}, accessed through their respective APIs with default decoding settings.

\section{Evaluations}
\label{sec:evaluations}

\begin{itemize}
    \item \textbf{RQ1: Autotuned Programming-Model Performance.} Under matched operator semantics and comparable implementation structures, how do the autotuned Triton and cuTile implementations compare with each other and with PyTorch baselines?
    \item \textbf{RQ2: Autotuning Sensitivity and Performance Headroom.} How much performance do the exposed configuration knobs recover beyond hand-selected defaults, and how much hardware headroom remains after tuning?
    \item \textbf{RQ3: Performance Gap Diagnosis.} Across TileBench, what compiler lowering and hardware execution behaviors explain the remaining gaps between aligned Triton and cuTile implementations?
    \item \textbf{RQ4: LLM Usability and Token Efficiency.} Under a matched generation and feedback protocol with a maximum budget of 10 refinement iterations and no backend autotuning, how do Triton and cuTile compare in correctness-gated speedup, token cost, and token efficiency for LLM-generated kernels?
\end{itemize}

\ref{subsec:rq1} reports the autotuned performance over the complete dtype and input-size sweeps. 
\ref{subsec:rq2} directly compares the default and autotuned results over the same sweep.
\ref{subsec:rq3} profiles the NCU-cataloged maximum input of each supported dtype using the autotuned winner selected for that exact case.
\ref{subsec:rq4} is evaluated separately: each model-backend pair is allowed up to ten refinement iterations and cannot invoke backend autotuners; once a backend satisfies the stopping criterion, it is frozen and incurs no additional token cost.
Unless otherwise stated, all aggregation above the per-case latency uses geometric means.

\subsection{RQ1: Autotuned Programming-Model Performance}
\label{subsec:rq1}

\begin{figure}
    \centering
    \includegraphics[width=0.8\linewidth]{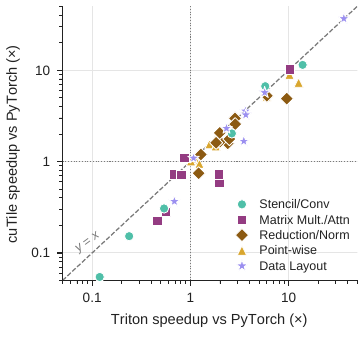}
    \caption{Triton and cuTile speedup over PyTorch across operators.}
    \label{fig:rq1}
\end{figure}

\textbf{Answer: Under autotuned configurations, Triton is stronger in aggregate and more robust across diverse workloads, while cuTile remains competitive and achieves its clearest advantages on regular, reuse-heavy tiled workloads, including Tensor-Core / TMA kernels.}

\Cref{fig:rq1} compares the operator-level autotuned speedups of Triton and cuTile over PyTorch.
Each point summarizes one operator using the geometric mean over all of its supported dtype and input-size cases.
We make three observations:

\noindent\textbf{(1) Both DSLs outperform PyTorch on most operators, but Triton is stronger overall.}
Triton achieves a geometric-mean speedup of $2.02\times$ over PyTorch, compared with $1.58\times$ for cuTile.
Triton is faster than PyTorch on $36/45$ operators, while cuTile is faster on $33/45$.
Their median speedups are $1.95\times$ and $1.58\times$, respectively, showing that the aggregate difference is not caused only by a few outliers.

\noindent\textbf{(2) The direct Triton and cuTile comparison is workload-dependent.}
Triton is faster on $37/45$ operators, while cuTile is faster on $8/45$.
However, the operator-level geometric-mean latency ratio between the two backends differs from parity by at most $5\%$ on $14/45$ operators and by at most $10\%$ on $17/45$, indicating that many head-to-head differences are small.
The clearest cuTile advantages occur when regular tile movement, larger Tensor Core tiles, or two-dimensional data reuse can amortize tile-staging costs.
Representative cases include \texttt{matmul\_fp32\_fp16\_fp8}, \texttt{flash\_attention}, and \texttt{jacobi\_stencil\_2d}.
In contrast, Triton's largest margins occur on runtime-computed indexing, dynamic scatter, fine-grained reductions, sparse traversal, and low-parallelism loops, as observed in \texttt{destindex}, \texttt{weight\_dequant}, \texttt{moe\_topk\_gating}, \texttt{flash\_decode}, and the convolution family.

\noindent\textbf{(3) Speedup over PyTorch does not always isolate a compiler-backend effect.}
PyTorch serves as a practical software baseline, but its internal execution path varies across operators.
For eager compositions such as \texttt{rope}, \texttt{dropout}, \texttt{kl\_divergence}, \texttt{fused\_activation}, the DSL advantage mainly comes from kernel fusion and avoiding intermediate tensors.
For reductions and indexing operators, such as \texttt{mean\_reduction}, \texttt{argmax}, \texttt{destindex}, the difference reflects a shape-specialized DSL kernel compared with general ATen implementation.
Conversely, PyTorch can be faster when it dispatches to highly optimized vendor implementation, including cuDNN convolution, cuBLAS GEMM, CUB radix sort, and warp-specialized attention kernels.

Additional distributions and category-level results are reported in \Cref{appsubsec:rq1_extension}.

\subsection{RQ2: Autotuning Sensitivity and Performance Headroom}
\label{subsec:rq2}

\begin{figure}
    \centering
    \includegraphics[width=0.8\linewidth]{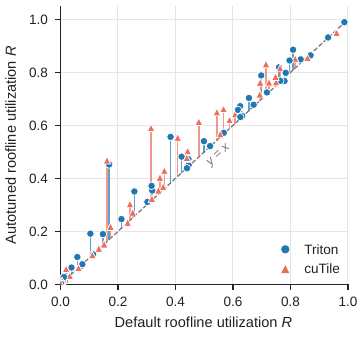}
    \caption{Default-to-autotuned roofline utilization across operators.}
    \label{fig:rq2}
\end{figure}

\textbf{Answer: Autotuning provides meaningful additional gains beyond hand-selected defaults for both backends, with a larger geometric-mean gain for cuTile, but substantial hardware headroom remains outside the exposed configuration spaces.}

\Cref{fig:rq2} compares each operator's geometric-mean default and autotuned roofline utilization.
The geometric-mean improvement is $1.18\times$ for Triton and $1.22\times$ for cuTile, while the median improvements are $1.07\times$ and $1.06\times$, respectively.
These gains are measured relative to manually selected defaults rather than random or deliberately weak configurations.
Autotuning therefore recovers meaningful performance from tile shape, thread organization, pipeline depth, and occupancy choices that are difficult to select uniformly across operators and dtypes.

The autotuned results should be interpreted as the best observed configurations within the declared search spaces.
Triton exposes block dimension, warp counts, pipeline stages, and operator-specific scheduling parameters, while cuTile exposes tile dimensions, occupancy hints, and operator-specific launch choices.
The search does not alter the source-level algorithm or explicitly explore alternative implementation structures, primitive choices, or lowering strategies.
RQ2 therefore measures \emph{configuration headroom}, not a global upper bound on either programming model.

After tuning, $7/45$ Triton implementations and $5/45$ cuTile implementations reach at least $80\%$ operator-level geometric-mean roofline utilization.
Only $2/45$ Triton implementations and $1/45$ cuTile implementation reach at least $90\%$.
Thus, autotuning closes part of the performance gap while also revealing substantial remaining \emph{backend headroom}.
These factors generally cannot be corrected by selecting another tile size from the same implementation.

\subsection{RQ3: Performance-Gap Diagnosis}
\label{subsec:rq3}

\textbf{Answer: The performance gaps between cuTile and Triton arise from recurring differences in index lowering, reduction granularity, state placement, and pipeline scheduling. cuTile's tensor core backend is not inherently weaker: it is highly effective on regular GEMMs, while its largest deficits occur when irregular indexing or fixed tile-lowering costs cannot be sufficiently amortized.}

\begin{figure}[t]
    \centering
    \includegraphics[width=\columnwidth]{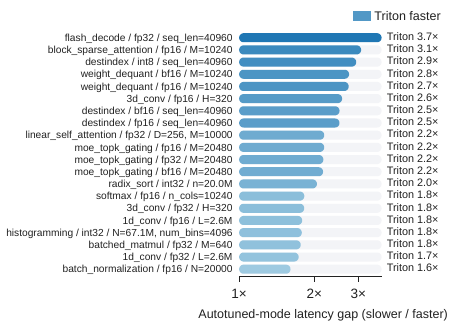}
    \caption{The 20 largest Triton-over-cuTile latency gaps under autotuned mode, evaluated at the NCU-profiled sweep-max input for each operator--dtype pair.
    Each bar represents one operator--dtype pair.
    The x-axis reports the ratio of the slower-backend latency to the faster-backend latency.}
    \label{fig:rq3-top-perf-gaps}
\end{figure}


\paragraph{Aligned Triton versus cuTile.}

\Cref{fig:rq3-top-perf-gaps} shows that all 20 of the largest sweep-max gaps favor Triton.
The largest are \texttt{flash\_decode}/FP32 at $3.72\times$, \texttt{block\_sparse\_attention}/FP16 at $3.08\times$, and \texttt{destindex}/INT8 at $2.94\times$.
The largest cuTile advantage, $1.55\times$ on FP32 \texttt{matmul\_fp32\_fp16\_fp8}, falls below the twentieth Triton-faster gap and is analyzed separately below.
Overall, the largest Triton advantages are concentrated in irregular indexing, dynamic scatter, small reductions, sparse or runtime-controlled loops, and cases where fixed tile-management costs are large relative to useful work.
The following two cases expose the contrast between irregular indexing and regular tiled computation.

\paragraph{Case A: index lowering in \texttt{weight\_dequant} and \texttt{destindex}.}

To investigate compiler-level performance discrepancies, we profile two memory-bound kernels—\textit{BF16 Weight Dequantization} (\texttt{weight\_dequant}) and \textit{Int8 Destination Indexing (\texttt{destindex})}—on B200. 
Despite their memory-bound nature, Triton substantially outperforms cuTile on both kernels.
NCU profiling reveals an artificial bottleneck inversion in cuTile: 
in BF16 \texttt{weight\_dequant}, cuTile saturates the Streaming Multiprocessor (SM) ($87.8\%$) and Arithmetic Logic Unit (ALU) ($89.7\%$) pipelines while Load/Store Unit (LSU) utilization drops to $12.8\%$, whereas Triton maintains a balanced profile ($28.2\%$ SM, $30.2\%$ ALU, and $23.1\%$ LSU).
A similar pattern occurs in Int8 \texttt{destindex}, where cuTile sustains $78.5\%$ ALU utilization versus $12.7\%$ in Triton. 
While Triton warps predominantly encounter \texttt{stall\_long\_scoreboard} (hiding memory latency), cuTile warps are severely congested by \texttt{stall\_not\_selected} and \texttt{stall\_math\_pipe\_throttle}, indicating that issue slots are choked by excessive integer arithmetic rather than memory stalls.

This compute saturation originates from severe instruction bloat during offset calculations rather than memory operations, and cuTile executes $6.2\times$ more total instructions in BF16 \texttt{weight\_dequant} and $17.3\times$ more in Int8 \texttt{destindex} (86.99M vs. 5.04M). 
SASS disassembly traces this to multi-dimensional indexing (e.g., \texttt{offsets // N} and \texttt{offsets \% N}): Triton lowers integer division/modulo via \texttt{arith.divsi}/\texttt{remsi} into constant reciprocal multiplications with magic constants (e.g., \texttt{0x66666667}). 
Conversely, cuTile lowers index arithmetic into defensive ``safe-division'' routines spanning much more SASS instructions (\texttt{ISETP}, \texttt{LOP3}, and \texttt{SEL}) to guard against signed inputs. 
Because tensor shapes are strictly non-negative compile-time constants (\texttt{tl.constexpr}), cuTile's index lowering leaves these value ranges unexploited, introducing entirely redundant arithmetic overhead.

Additionally, cuTile exhibits degraded memory coalescing in the \texttt{destindex} kernel, achieving only $1.78$ sectors per load request ($\text{\texttt{sectors\_sum}} / \text{\texttt{requests\_sum}}$) compared to Triton's $8.50$. 
This disparity shows that Triton vectorizes contiguous thread accesses into wide vectorized transactions, whereas cuTile fragments loads into disjoint, uncoalesced requests. In summary, the compounding effects of defensive arithmetic instruction bloat and memory transaction fragmentation fully account for cuTile's performance degradation.

\paragraph{Case B: regular tiled computation in \texttt{matmul\_fp32\_fp16\_fp8}.}
Matrix multiplication provides a control case because the irregular indexing chain disappears.
Both implementations use grouped tiled GEMM, FP32 accumulation, Tensor-Core MMA, and TMA-backed memory movement.
The performance difference therefore cannot be explained by cuTile using TMA while Triton does not.

NCU instead shows that cuTile keeps the Tensor pipeline active for approximately \(79\%\)--\(81\%\) of cycles, compared with roughly \(52\%\)--\(69\%\) for Triton.
At \(M=N=4096\) and \(K=20480\), cuTile is \(1.55\times\) faster in FP32, \(1.12\times\) faster in FP16, and \(1.22\times\) faster in FP8.
The FP32 difference is largest because cuTile sustains a \(256\times256\times64\) tile, while Triton's TMA store staging and multi-stage buffers constrain its selected configuration to \(128\times128\times32\).
The smaller Triton tile nearly doubles the measured TMA load volume, \(21.5\) GB compared with \(10.7\) GB, while leaving scheduling and data-movement overhead amortized over less MMA work.

\subsection{RQ4: LLM Usability and Token Efficiency}
\label{subsec:rq4}

\begin{figure}
    \centering
    \includegraphics[width=0.8\linewidth]{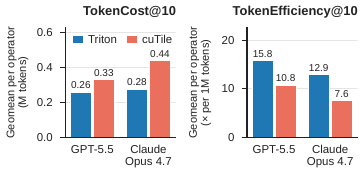}
    \caption{LLM token cost and efficiency by backend.}
    \label{fig:rq4}
\end{figure}

\textbf{Answer: Under the same maximum 10-iteration protocol, Triton is a more effective and token-efficient target for both evaluated LLMs. It produces verify-clean kernels for more operators, reaches higher correctness-gated speedup, and consumes fewer tokens than cuTile.}

\Cref{fig:rq4} reports the geometric-mean \(\mathrm{TokenCost@10}\) and \(\mathrm{TokenEfficiency@10}\) for each model-backend combination.
We additionally report the geometric-mean \(\mathrm{BestSpeedup@10}\) over operators with at least one verify-clean result to separate final performance from token expenditure.

For GPT-5.5, Triton produces a faster-than-PyTorch verify-clean kernel on \(38/45\) operators.
cuTile does so on \(36/43\) operators with at least one verify-clean result, with two operators failing to produce a verify-clean cuTile implementation.
The geometric-mean \(\mathrm{BestSpeedup@10}\) is \(4.12\times\) for Triton and \(3.46\times\) for cuTile.
The corresponding token costs are \(0.260\)M and \(0.331\)M tokens, while token efficiency is \(15.83\times\) and \(10.79\times\) per million tokens, respectively.

For Claude Opus 4.7, Triton produces a faster-than-PyTorch verify-clean kernel on \(38/45\) operators.
cuTile does so on \(33/41\) operators with at least one verify-clean result, with four operators failing to produce a verify-clean cuTile implementation.
The geometric-mean \(\mathrm{BestSpeedup@10}\) is \(3.57\times\) for Triton and \(3.16\times\) for cuTile.
The token cost increases from \(0.277\)M tokens on Triton to \(0.441\)M on cuTile, while token efficiency decreases from \(12.90\times\) to \(7.61\times\) per million tokens.

The backend effect is consistent across both models.
Relative to cuTile, Triton improves token efficiency by approximately \(47\%\) for GPT-5.5 and \(70\%\) for Claude Opus 4.7.
The difference comes from both sides of the metric: Triton more often reaches a stronger verify-clean implementation, and the refinement process consumes fewer tokens while doing so.
The model effect is less uniform.
GPT-5.5 achieves the strongest Triton result, while the differences between the two models are smaller than the corresponding Triton and cuTile gaps.

These results establish an empirical usability difference under the TileBench protocol, but they do not isolate its cause.
Possible explanations include differences in API expressibility, available primitives, compiler constraints, documentation, examples, and model exposure to each DSL.
Because the experiment does not separately control these factors, we treat reduced model familiarity with cuTile as a possible explanation rather than a demonstrated causal mechanism.

\section{Conclusions}

We introduce TileBench, a controlled and reproducible benchmark for comparing tile-based GPU programming models and studying tile-centric DSLs. Our results show that both Triton and cuTile substantially outperform PyTorch, but neither abstraction is uniformly superior: Triton is more robust on irregular, streaming, and bandwidth-bound operators, while cuTile excels when workloads match static tile abstractions, Tensor Core computation, and reusable tile movement. Autotuning yields measurable gains, yet most kernels remain far from roofline limits, suggesting that backend lowering, memory staging, instruction selection, and shared-memory behavior matter more than tile-size search alone. The LLM-generated track further shows that programming-model design affects both final performance and the token cost of reaching correct, efficient kernels.

\section*{Limitations}

TileBench currently targets a single NVIDIA B200 GPU and two tile-based DSLs.
Its roofline metrics use analytical FLOP/HBM-byte formulas, so they provide a standardized hardware-aware summary rather than a replacement for full NCU-counter analysis.

\section*{Acknowledgments}
This material is based upon work supported by the National Science Foundation under Award Nos.~2549919 and 2554012, and a donation from NVIDIA.
We thank the anonymous reviewers for their valuable feedback. We acknowledge the use of Claude Code and OpenAI Codex for language editing and code implementation. No AI tools were used to generate research ideas, analyses, or experimental results. All code and text were reviewed, tested, and validated by the authors, who take full responsibility for the contents of this work.

\bibliography{reference}

\clearpage
\appendix

\crefalias{section}{appendix}
\crefalias{subsection}{appendix}  
\crefalias{subsubsection}{appendix}
\Crefname{appendix}{Appendix}{Appendices}
\crefname{appendix}{Appendix}{Appendices}

\section*{Appendix}
\label{sec:appendix}

\section{Benchmark Information}
\label{appsec:benchamrk_information}

\subsection{Full Benchmark List}
\label{appsubsec:full_benchmark_list}

\begin{strip}
\centering
\small

\captionof{table}{TritonBench-derived operators in TileBench.}
\label{tab:operators_tritonbench}

\begin{tabularx}{\textwidth}{@{}p{3.8cm} p{3.2cm} X@{}}
\toprule
\textbf{Operator Name} &
\textbf{Category} &
\textbf{Explanation} \\
\midrule

flash-attention
& Matrix Mult./Attn.
& Standard prefill/dense attention.\\

flash-decode
& Matrix Mult./Attn.
& Flash decoding attention.\\

block-sparse-attn
& Matrix Mult./Attn.
& Block-sparse attention.\\

softmax
& Reduction/Norm.
& Standard Softmax.\\

rope-embedding
& Point-wise
& Positional embedding.\\

matmul\_fp32\_fp16\_fp8
& Matrix Mult./Attn.
& Standard GEMM (supports FP16 \& FP8).\\

matmul-int8
& Matrix Mult./Attn.
& Quantized GEMM.\\

streamk-matmul
& Matrix Mult./Attn.
& Stream-K scheduling optimization.\\

layernorm
& Reduction/Norm.
& Standard layer normalization.\\

rmsnorm
& Reduction/Norm.
& RMS normalization.\\

vector-addition
& Point-wise
& Binary addition.\\

mul2
& Point-wise
& Elementwise multiplication.\\

relu
& Point-wise
& Activation function.\\

destindex
& Data Layout
& Indexing/scatter/gather operations.\\

quantize-global
& Point-wise
& Packing values.\\

dequantize-rowwise
& Point-wise
& Unpacking values.\\

cross-entropy
& Reduction/Norm.
& Elementwise loss calculation.\\

matrix-transpose
& Data Layout
& Transpose / permute.\\

dropout
& Point-wise
& Random mask generation + apply.\\

swiglu
& Point-wise
& Complex gated activation.\\

kl-divergence
& Reduction/Norm.
& Divergence metric calculation.\\

fused-activation
& Point-wise
& Generic fused kernel container.\\

mean-reduction
& Reduction/Norm.
& Sum/mean reduction.\\

argmax
& Reduction/Norm.
& Index finding.\\

l2-norm
& Reduction/Norm.
& Norm calculation.\\

2d-conv
& Stencil/Conv.
& 2D convolution.\\

\bottomrule
\end{tabularx}

\vspace{0.8\baselineskip}


\captionof{table}{LeetGPU-derived operators in TileBench.}
\label{tab:operators_leetgpu}

\begin{tabularx}{\textwidth}{@{}p{3.8cm} p{3.2cm} X@{}}
\toprule
\textbf{Operator Name} &
\textbf{Category} &
\textbf{Explanation} \\
\midrule

top-k-selection
& Data Layout
& Select the top-$K$ largest elements from an array.\\

histogramming
& Reduction/Norm.
& Count occurrences of each value into histogram bins.\\

linear-self-attn
& Matrix Mult./Attn.
& Compute linear attention using kernel feature maps.\\

jacobi-stencil-2d
& Stencil/Conv.
& Jacobi stencil 2D.\\

bitonic-sort
& Data Layout
& Sort an array of floats in ascending order.\\

radix-sort
& Data Layout
& Sort 32-bit unsigned integers in ascending order via radix sort.\\

batched-matmul
& Matrix Mult./Attn.
& Batched matrix multiplication.\\

batch-normalization
& Reduction/Norm.
& Batch normalization.\\

2d-max-pooling
& Stencil/Conv.
& 2D max pooling.\\

gaussian-blur
& Stencil/Conv.
& Apply a Gaussian blur filter via 2D convolution.\\

1d-conv
& Stencil/Conv.
& 1D convolution.\\

3d-conv
& Stencil/Conv.
& 3D convolution.\\

matrix-copy
& Data Layout
& Copy an $N\times N$ matrix element-wise.\\

reverse-array
& Data Layout
& Reverse an array in-place.\\

interleave
& Data Layout
& Interleave two float arrays element-wise.\\

weight-dequant
& Point-wise
& Dequantize a tiled weight matrix using per-tile scale factors.\\

moe-topk-gating
& Reduction/Norm.
& Select the top-$k$ experts per token and compute softmax gating.\\

leaky-relu
& Point-wise
& Leaky ReLU activation.\\

sigmoid
& Point-wise
& Sigmoid activation.\\

\bottomrule
\end{tabularx}

\end{strip}    


\subsection{Operator Category}

\begin{table*}[!t]
\centering
\small
\begin{tabular}{lcl}
\toprule
Category & \#Ops & Representative operators \\
\midrule
Point-wise &
12 &
\texttt{vector\_add}, \texttt{relu}, \texttt{swiglu}, \texttt{weight\_dequant} \\
Reduction/Normalization &
11 &
\texttt{softmax}, \texttt{layernorm}, \texttt{argmax}, \texttt{moe\_topk\_gating} \\
Matrix Multiplication/Attention &
8 &
\texttt{matmul}, \texttt{matmul\_int8}, \texttt{flash\_attention}, \texttt{linear\_self\_attention} \\
Stencil/Convolution &
6 &
\texttt{1d\_conv}, \texttt{3d\_conv}, \texttt{gaussian\_blur}, \texttt{jacobi\_stencil\_2d} \\
Data Layout &
8 &
\texttt{matrix\_copy}, \texttt{matrix\_transpose}, \texttt{bitonic\_sort}, \texttt{radix\_sort} \\
\midrule
Total & 45 & \\
\bottomrule
\end{tabular}
\caption{TileBench operator categories. TileBench contains 45 operators:
26 adapted from TritonBench and 19 adapted from LeetGPU.}
\label{tab:tilebench_categories}
\end{table*}



\subsection{\texttt{Config.yaml} Example}
\label{appsubsec:config.yaml_example}

Each operator ships with a \texttt{config.yaml} file declaring the case grid the harness will run (the cross product of input sizes
and dtypes) and the verification tolerances applied to the generated kernel's output against the PyTorch reference.
The configuration is per-operator (not global) because different operators have meaningfully different shape-sensitivity profiles:
matmuls are exercised at sizes where tensor-core paths kick in,
reductions at sizes where occupancy starts saturating, and elementwise ops at sizes that span memory-bound and compute-bound regimes. 
The verify-tolerance defaults follow standard mixed-precision practice;
per-operator overrides are used where the reference itself is non-associative across reduction order, in which case the tolerance is widened to absorb floating-point reassociation error.

\subsection{B200 JSON Information}
\label{appsubsec:B200_JSON_Information}

We use per-GPU dense peak compute ceilings derived from the NVIDIA HGX B200 / DGX B200 specifications, excluding 2:4 sparsity. 
For FP32 dot/MMA workloads, we use the TF32 Tensor Core dense ceiling because Triton \texttt{tl.dot} uses TF32 by default for FP32 inputs. 
The peak bandwidth value, 6539.4 GB/s, is measured on our B200 system rather than taken from the datasheet HBM bandwidth.

\begin{figure}[!t]
\centering
\begin{lstlisting}
  "gpu": "NVIDIA B200",
  "peak_bw_GBs": 6539.4,
  "peak_tflops": {
    "fp64": 37,
    "fp32": 1100,
    "tf32": 1100,
    "fp16": 2250,
    "bf16": 2250,
    "fp8_e4m3fn": 4500,
    "fp8_e5m2": 4500,
    "int8": 4500
  },
\end{lstlisting}
\caption{B200.json Example}
\label{listing:b200}
\end{figure}

\begin{figure}[!t]
\centering
\begin{lstlisting}
benchmark:
  warmup: 20
  repeat: 100
  use_cuda_graph: true
  flush_l2: true
  autotune: false
case_defaults:
  M: 2048
case_grid:
  N: [20480]
  dtype: ["fp16", "fp32"]
metrics:
  flops_expr: "M * N"
  bytes_expr: "M * N * dtype_size"
  plots:
    - latency_ms
    - bandwidth_GBs
    - speedup
\end{lstlisting}
\caption{\texttt{argmax} (\texttt{argmax.yaml}) Example}
\label{listing:argmax_config}
\end{figure}

\section{Iterative LLM Code-Generation Prompts}
\label{appsec:iterative}

This appendix documents the prompts used in the iterative LLM code-generation track described in \Cref{subsec:iteration}.
We show the iteration-0 and iteration-$N$ prompt templates, followed by the building blocks they reference: the TileBench framework guide, the per-backend API references, the operator problem description, the PyTorch reference implementation, and the strict output-format instructions.
The operator's \texttt{config.yaml} is shown separately in \Cref{appsubsec:config.yaml_example}.
Throughout this section we use \texttt{vector\_add} as the running example operator; for every other operator the same skeleton is filled in with the corresponding operator-specific files.

\subsection{Iteration 0 Prompt Template}
\label{appsubsec:iter0_prompt}

The iteration-0 prompt is assembled by concatenating the components below in the order shown.
The model receives no prior context other than the system prompt; everything operator-specific is fed as the user message.

\begin{promptbox}[Iteration 0 Prompt Template]
# Task: implement operator `<op>` for TileBench

Follow the framework conventions below. Then implement the operator
described. You will emit `impl_triton.py`, `impl_cutile.py`.

---

# TileBench Framework Conventions
<<< framework guide -- Cref{appsubsec:framework_guide} >>>

---

# Triton (triton / triton.language) API Reference
<<< Triton API reference -- Cref{appsubsec:api_reference} >>>

---

# cuTile (cuda.tile) API Reference
<<< cuTile API reference -- Cref{appsubsec:api_reference} >>>

---

# Operator description: `<op>`
<<< natural-language problem description --
    Cref{appsubsec:problem_description} >>>

---

# `config.yaml` for `<op>`
<<< config.yaml -- Cref{appsubsec:config.yaml_example} >>>

---

# PyTorch reference: `impl_torch.py`
<<< impl_torch.py -- Cref{appsubsec:torch_reference} >>>

---

<<< output-format instructions --
    Cref{appsubsec:output_format} >>>
\end{promptbox}

\subsection{Iteration $N$ Prompt Template ($N>0$)}
\label{appsubsec:iterN_prompt}

For iterations after iteration~0, the prompt is rebuilt from scratch each iteration rather than appended to a running chat.
It still ends with the framework guide, API references, problem description, \texttt{config.yaml}, PyTorch reference, and output-format instructions (identical to iteration~0), but is prefixed with a trajectory table, the previous iteration's source files, and, when the previous iteration regressed or failed verification, the best verify-clean source files seen so far as a rebuild reference.

\begin{promptbox}[Iteration $N$ Prompt Template ($N>0$)]
# Task: improve operator `<op>` for TileBench (iteration N)

This is iteration N of the refinement loop. Review the trajectory
and feedback below. If the latest iteration regressed or failed
verification, consider rebuilding from the best verify-clean implementation seen so far. 
Then re-emit the implementation file(s) for the backend(s) that remain active.

## Iteration trajectory so far

<<< per-iteration backend configuration, stop_score,
    speedup_vs_torch, and verification status >>>

<<< best verify-clean iteration for each backend and
    regression warnings, when applicable >>>

## Feedback from iteration N-1

<<< compile/timeout errors, verification failures,
    per-backend performance summary, and worst-first
    roofline feedback >>>

---

## Previous generated implementations

```python
<previous impl_triton.py source>
```

```python
<previous impl_cutile.py source>
```

---

## Best verify-clean implementation

<<< included only when the latest implementation regressed
    or failed verification and an earlier verify-clean
    implementation is available >>>

```python
<best verify-clean impl_<backend>.py source>
```

---

# Reminders: TileBench Framework Conventions
<<< framework guide -- \Cref{appsubsec:framework_guide} >>>

---

# Triton (triton / triton.language) API Reference
<<< Triton API reference -- \Cref{appsubsec:api_reference} >>>

---

# cuTile (cuda.tile) API Reference
<<< cuTile API reference -- \Cref{appsubsec:api_reference} >>>

---

# Operator description: `<op>`
<<< problem description -- \Cref{appsubsec:problem_description} >>>

---

# `config.yaml` for `<op>`
<<< config.yaml -- \Cref{appsubsec:config.yaml_example} >>>

---

# PyTorch reference: `impl_torch.py`
<<< impl_torch.py -- \Cref{appsubsec:torch_reference} >>>

---

<<< output-format instructions --
    \Cref{appsubsec:output_format} >>>
\end{promptbox}

\subsection{TileBench Framework Guide}
\label{appsubsec:framework_guide}

The framework guide specifies the file structure each implementation must produce, the \texttt{run(*args, **kwargs)} and \texttt{get\_last\_config()} contract, the forbidden patterns (\texttt{triton.autotune}, \texttt{cutileAutotuner}, delegation to PyTorch / cuDNN / cuBLAS, output caching), recommended tile-size bounds for B200, multi-dtype handling, and a pre-return checklist.
An abridged transcription is shown below: markdown emphasis and decorative emoji from the source file have been stripped, but section headings, rules, code templates, and numeric bounds are reproduced as in the original.

\begin{promptbox}[TileBench Framework Guide (\texttt{framework\_guide.md})]
# TileBench Framework Guide

Read this carefully. Your generated kernels MUST follow this contract
or the benchmark harness will fail to load / run / evaluate them.

## Files you must produce

For operator `<op>`, write exactly two files:

1. `impl_triton.py` -- Triton kernel + Python `run()` wrapper
2. `impl_cutile.py` -- cuTile kernel + Python `run()` wrapper

You do NOT generate `impl_torch.py` -- it is copied from the
human-written reference implementation. Your `run()` function's
signature MUST match `impl_torch.run()` exactly (same positional
args, same keyword args).

## Required exports (both impls)

<<< `run(...)` and `get_last_config()` interface definitions >>>

## No autotune -- you pick one configuration per iteration

The harness does NOT call autotune. 
Each active backend runs with a single hard-coded configuration that YOU choose.
The refinement loop, with a maximum budget of 10 iterations per backend, is the search mechanism: at each executed iteration, the model sees the previous iteration's (config -> roofline_pct, latency_ms, speedup_vs_torch) and proposes the next configuration.

The following are FORBIDDEN in generated code and will be rejected
at load:

- `triton.autotune(...)` decorator
- `from core.cutile_autotune import CutileAutotuner`
- `ct.tune.exhaustive_search` / `ct_experimental.autotune_launch`
- Any other in-kernel autotune-style search

## Triton template (no autotune)

<<< Triton template and configuration-recording rules >>>

## cuTile template (no autotune)

<<< cuTile template and backend-specific implementation rules >>>

## Choosing tile / BLOCK sizes

<<< B200-specific tile-size bounds and recommended starting
    configurations >>>

## Multi-dtype support

If `config.yaml`'s `case_grid.dtype` lists multiple dtypes (e.g.
["fp16", "bf16", "fp32"]), the `run()` function must work for ALL
of them in a single `run()` call with ONE configuration shared
across all dtypes. Use a single dtype-polymorphic kernel:

<<< backend-specific multi-dtype implementation guidance >>>

## Hardware constraints to remember (B200, sm_100)

<<< B200-specific TMA, Tensor Core, and tile-shape constraints >>>

## FORBIDDEN: delegating computation to PyTorch / cuDNN / cuBLAS

The `run()` function in `impl_triton.py` MUST perform the
operator's computation through the `@triton.jit` kernel, and
`run()` in `impl_cutile.py` MUST perform it through the
`@ct.kernel`. The following are NOT allowed inside `run()`:

<<< prohibited delegation patterns and permitted non-compute
    PyTorch operations >>>

## FORBIDDEN: caching outputs across `run()` calls

Do NOT add a Python-level cache that returns a previously-computed
output when the inputs look identical.

<<< anti-cache constraints and harness checks >>>

## Common pitfalls (the harness rejects these)

<<< framework validation and common implementation pitfalls >>>

## What the engine does with generated code

<<< per-case input generation, verification, timing, and
    configuration-feedback procedure >>>

The refinement loop allows up to 10 iterations per backend.
Triton and cuTile are tracked independently.
After each executed iteration, the engine evaluates verification and roofline utilization.
A backend is frozen once its implementation passes verification and reaches at least 80% roofline utilization.
Otherwise, that backend remains active until the 10-iteration budget is exhausted.

Frozen backends are skipped in subsequent iterations and incur no additional token cost.
When refinement ends, the best verification-passing iteration among the executed iterations is promoted as the final result.
The `stop_score` recorded in the trajectory corresponds to the roofline utilization criterion used for stopping and is included in the feedback for subsequent iterations.

## TL;DR checklist before returning code

<<< pre-return checklist >>>
\end{promptbox}

\subsection{Triton and cuTile API References}
\label{appsubsec:api_reference}

For each active backend, the prompt prepends an authoritative API reference that scopes the model to the exact library versions used by the harness (Triton 3.6.0 and the cuTile pip packages shipped with TileBench).
Each reference is a long document; reproducing either in full would dominate the appendix, so we describe their structure and show a representative excerpt.

\paragraph{Structure.}
Each API reference is organized as numbered topical sections that, at a high level, cover (i) import boilerplate and data types, (ii) the kernel programming model and tile creation, (iii) memory operations, (iv) compute primitives (arithmetic, reductions, matrix multiply), (v) control flow and synchronization, (vi) compiler hints, (vii) the autotune pattern (explicitly disabled in this pipeline; see the preamble below), (viii) the TileBench \texttt{run()} / \texttt{get\_last\_config()} contract and common kernel patterns, and (ix) critical constraints, gotchas, and a quick-reference cheatsheet.
The exact section list differs between the two references because Triton and cuTile expose different abstractions.

\paragraph{Preamble (Triton).}
Each reference is prefixed by a short preamble that anchors the model to the installed library version and explicitly invalidates the reference's autotune content (since the LLM track forbids autotune).
The Triton preamble is:

\begin{promptbox}[Triton API Reference Preamble]
Treat the API reference below as authoritative for the Triton version
installed in this repo (3.6.0); do NOT use APIs from later versions
you may have seen in training data. Every `tl.*` / `triton.*` symbol
you use must appear in this reference.

IMPORTANT -- autotune sections of this reference do NOT apply. Use
the reference for kernel-body syntax (tl.load, tl.store, tl.dot,
masking, make_block_ptr, make_tensor_descriptor, etc.); ignore
everything about cfg search / autotune wrappers.
\end{promptbox}

The cuTile preamble is analogous.

\paragraph{Section excerpt.}
As a concrete example of the section-level granularity, the cuTile reference's memory-operations section opens as shown below.
Each entry gives the function signature, a parameter list, and a few representative call examples; per-section content varies in length but follows this shape.

\begin{promptbox}[cuTile API Reference: \S 6 Memory Operations (Excerpt)]
## 6. Memory Operations

### 6.1 ct.load()

Signature:
    tile = ct.load(array, index, shape, *,
                   order="C",
                   padding_mode=PaddingMode.UNDETERMINED,
                   latency=None,
                   allow_tma=None)

Parameters:
    array        : Array
                   Source array (torch tensor)
    index        : tuple[int | Tile, ...]
                   Tile-space coordinates
    shape        : tuple[const int, ...]
                   Tile shape (all dims must be power of 2);
                   () for 0D scalar tile
    order        : "C" / "F" / tuple
                   Axis mapping: "C"=identity, "F"=reversed,
                   tuple=explicit permutation
    padding_mode : PaddingMode
                   OOB element handling
    latency      : const int
                   DRAM traffic hint, 1 (low) to 10 (high)
    allow_tma    : const bool
                   Enable/disable TMA (default True)

Examples:
    tile   = ct.load(x, index=(bid,), shape=(TILE,))             # 1D
    tile   = ct.load(x, index=(row, j), shape=(1, TILE),
                     padding_mode=ct.PaddingMode.ZERO)            # 2D padded
    tile   = ct.load(Q, index=(b, h, i, 0), shape=(1,1,M,D))     # 4D
    tile   = ct.load(K, index=(b, h, 0, j), shape=(1,1,D,N),
                     order=(0,1,3,2))                             # transposed
    scalar = ct.load(x, index=(i,), shape=())                     # 0D scalar

### 6.2 ct.store()

Signature:
    ct.store(array, index, tile, *,
             order="C",
             latency=None,
             allow_tma=None)

Parameters:
    array     : Array
                Destination array
    index     : tuple[int | Tile, ...]
                Tile-space coordinates
    tile      : Tile or scalar
                Data to store
    order     : "C" / "F" / tuple
                Axis mapping
    latency   : const int
                Latency hint
    allow_tma : const bool
                Enable/disable TMA

OOB writes are silently ignored (no-op).
\end{promptbox}

\subsection{Operator Problem Description}
\label{appsubsec:problem_description}

Each operator has a natural-language problem description at \texttt{benchmarks/\allowbreak problems/\allowbreak current/\allowbreak <op>\_current.md} that gives the mathematical definition, the input contract, the output contract, and a small worked example.
Below we show the file for the \texttt{vector\_add} operator as a representative example.

\begin{promptbox}[Problem Description (\texttt{vector\_add\_current.md})]
# Vector Addition

Implement a program that performs element-wise addition of two 1D
arrays of the same length.

The input consists of two arrays:

- `x`: A 1D array.
- `y`: A 1D array of the same length as `x`.

The output should be written to the `output` array, which has the
same length as the inputs.

The operation is defined mathematically as:

    output[i] = x[i] + y[i]

where i ranges from 0 to n-1.

## Implementation Requirements

- Use only native features (external libraries are not permitted)
- The `solve` function signature must remain unchanged
- The final result must be stored in the array `output`

## Example 1

    Input:  x = [1, 2, 3, 4], y = [5, 6, 7, 8]
    Output: [6, 8, 10, 12]

## Example 2

    Input:  x = [-1.5, 2.0, 0.0], y = [0.5, -2.0, 3.0]
    Output: [-1.0, 0.0, 3.0]

## Constraints

- 1 <= n
- `x` and `y` have the same length
\end{promptbox}

\subsection{PyTorch Reference Implementation}
\label{appsubsec:torch_reference}

The PyTorch reference (\texttt{impl\_torch.py}) is included verbatim in the prompt.
It serves both as a precise specification of the operator's input/output contract and as the correctness oracle the generated kernels are validated against.
For \texttt{vector\_add} the reference is:

\begin{promptbox}[PyTorch Reference (\texttt{vector\_add/impl\_torch.py})]
def run(x, y):
    return x + y
\end{promptbox}

The harness checks that the generated kernels' \texttt{run()} has the same positional signature as the reference, and that its output matches the reference within the tolerances declared in \texttt{config.yaml}'s \texttt{verify:} block (or per-dtype defaults when absent).

\subsection{Output Format Instructions}
\label{appsubsec:output_format}

The final block of every prompt enforces a strict, machine-parseable output format.
Generated text is parsed by a regular expression that matches \texttt{```python title="impl\_<backend>.py"} fenced code blocks; anything outside the requested blocks is discarded.
At each executed iteration, implementation files are emitted for the backend(s) that remain active, following the instructions below:

\begin{promptbox}[Output Format Instructions]
## Output format (STRICT)

Return the fenced code block(s) for the active backend(s), in the order shown below, using the corresponding title(s):

    ```python title="impl_triton.py"
    # full Python file content here
    ```

    ```python title="impl_cutile.py"
    # full Python file content here
    ```
Emit ONLY the implementation file(s) for the active backend(s).
No other code blocks. 
No prose between or after the blocks beyond a 1-2 sentence summary of your approach. Do NOT include test code, do NOT include PyTorch reference code (that file is provided by the framework).

\end{promptbox}

\section{Additional Evaluation Metrics}
\label{appsec:additional_evaluation_metrics}

\subsection{Evaluation Modes and Aggregation Conventions}
\label{appsubsec:evaluate_modes}

\paragraph{Per-case timing.}
For operator $o$, backend $b$, dtype $d$, and input case $c$, the benchmark records latency $T_{o,b,d,c}$. This is the arithmetic mean of 100 timed launches following 20 warmup launches. CUDA graph replay is enabled, and the L2 cache is evicted before each measured launch outside the timed Proton scope. The complete operator \texttt{run()} is timed as one unit, so a multi-kernel implementation retains any intended producer-to-consumer cache reuse inside the operator boundary.

\paragraph{Geometric-mean aggregation.}
Every paper-level aggregate above the per-case latency uses a geometric mean. The per-case speedup over PyTorch is
\begin{equation}
S_{o,b,d,c}
=
\frac{T_{o,\mathrm{torch},d,c}}{T_{o,b,d,c}}.
\end{equation}
For the complete sweep of operator $o$, we assign equal weight to every valid dtype and case pair:
\begin{equation}
S_{o,b}
=
\operatorname{GM}_{(d,c)\in\mathcal{V}_o}
S_{o,b,d,c},
\end{equation}
where $\mathcal{V}_o$ is the set of valid cases for operator $o$. The suite-level result gives equal weight to each of the 45 operators:
\begin{equation}
S_b^{\mathrm{suite}}
=
\operatorname{GM}_{o\in\mathcal{O}} S_{o,b}.
\end{equation}
Category results use the same construction over operators in the category. The direct operator-level Triton and cuTile ratio is
\begin{equation}
Q_o
=
\frac{S_{o,\mathrm{Triton}}}{S_{o,\mathrm{cuTile}}}.
\end{equation}
Thus, $Q_o>1$ favors Triton and $Q_o<1$ favors cuTile. Because the PyTorch factor cancels case by case, $Q_o$ is also the geometric mean of $T_{o,\mathrm{cuTile},d,c}/T_{o,\mathrm{Triton},d,c}$ over the complete sweep. Medians reported in RQ1 and RQ2 are distribution statistics over the 45 operator-level values and are not used as averages.

\paragraph{Roofline aggregation and tuning gain.}
Let $R_{o,b,d,c}^{m}$ denote the roofline utilization of mode $m\in\{\mathrm{default},\mathrm{autotuned}\}$. The operator-level value plotted in the paper is
\begin{equation}
\bar{R}_{o,b}^{m}
=
\operatorname{GM}_{(d,c)\in\mathcal{V}_o} R_{o,b,d,c}^{m}.
\end{equation}
The operator-level autotuning gain is
\begin{equation}
G_{o,b}
=
\operatorname{GM}_{(d,c)\in\mathcal{V}_o}
\frac{R_{o,b,d,c}^{\mathrm{autotuned}}}
     {R_{o,b,d,c}^{\mathrm{default}}},
\end{equation}
and the suite-level gain is $\operatorname{GM}_{o}G_{o,b}$. Threshold counts such as $R\geq 0.8$ are evaluated on $\bar{R}_{o,b}^{\mathrm{autotuned}}$.

\paragraph{RQ3 case selection.}
RQ3 does not average NCU counters across the sweep. 
For each supported operator--dtype pair, we select the maximum case recorded in the NCU catalogue and profile the exact autotuned winner used for that case. Performance ratios remain based on the corresponding Proton/CSV latency. NCU counters explain the execution mechanism but do not replace the benchmark latency.

\paragraph{RQ4 aggregation.}
For the LLM track, the correctness-gated speedup of an executed iteration is the PyTorch-to-generated-kernel latency ratio of the representative large verify-clean case saved by the evaluation pipeline. 
It is a single-case ratio rather than an average over dtypes. 
\(\mathrm{BestSpeedup@10}\) is the maximum verify-clean speedup observed before the run ends or the backend freezes. 
\(\mathrm{TokenCost@10}\) sums the tokens consumed by the executed iterations for that backend; skipped iterations after freezing contribute neither tokens nor speedup. 
\(\mathrm{TokenEfficiency@10}\) divides the best speedup by token cost in millions.

\subsection{NCU Profiling Protocol and Kernel Attribution}

The systematic Triton and cuTile profiling set contains 220 profiles: 110 supported operator--dtype pairs for Triton and the same 110 pairs for cuTile. 
Each profile uses the NCU-catalogued maximum input and the same autotuned winning configuration evaluated by the main benchmark. 
PyTorch is profiled selectively when its internal dispatch path is needed to interpret the software-baseline comparison, for example cuDNN convolution, cuBLAS GEMM, CUB radix sort, compiled \texttt{flex\_attention}, or a generic ATen reduction.

\paragraph{Latency source.}
All operator latency and speedup values are obtained from Proton measurements. NCU is used only for profiling and diagnosing performance phenomena, and its reported kernel duration is not used in our performance results.

\paragraph{Multi-kernel attribution.}
For a single-kernel implementation, the profile directly corresponds to the measured operator. 
For a multi-kernel implementation, the CSV latency covers the full kernel sequence, while counter-level observations are attributed to the stage that generates them. 
Examples include the two \texttt{destindex} scatter kernels; the five \texttt{linear\_self\_attention} launches grouped into activation, KV-GEMM, column-reduction, and output-GEMM stages; and the repeated histogram, scan, and scatter stages of \texttt{radix\_sort}. 


\paragraph{Diagnostic procedure.}
For each profiled case, we analyze the performance gap as follows:
\begin{enumerate}
    \item Inspect the source implementation, including the algorithm, task decomposition, precision path, and selected autotuned configuration;
    \item Inspect the generated PTX, SASS, and source-correlated hotspots;
    \item Examine dynamic instruction counts, memory requests and sectors, TMA and Tensor Core usage, register and shared-memory usage, occupancy, pipeline activity, and stall behavior;
    \item Relate these profiling observations to the corresponding Proton-measured latency.
\end{enumerate}

Higher-level compiler IRs were not materialized in the finalized artifacts used for this analysis, and no compilation was rerun during validation.
We therefore do not quote TTIR, TTGIR, or tile IR.
Compiler-lowering claims are instead based on PTX, SASS, and source-correlated evidence preserved in the NCU reports.
Opcode-family counts are extracted from NCU source tables.
When source correlation undercounts instructions associated with a source line, the counter-level \texttt{smsp\_\_inst\_executed.sum} total remains authoritative, and the per-family executed count is treated as a lower bound.

A stall metric alone is not treated as causal evidence.
For example, a long-scoreboard sample on an MMA line may indicate that the MMA is waiting for a preceding gather or staging chain rather than that Tensor Core arithmetic is intrinsically slow.
We therefore interpret stalls together with instruction mix, dependency producers, resource use, memory transactions, and measured latency.

\paragraph{Scope of the diagnosis.}
The profiles describe the current source implementations, compiler versions, autotuned winners, and B200 architecture.
They support recurring mechanisms across the suite, but they do not imply that every occurrence of a given source primitive must produce the same hardware path.
The generated instruction sequence remains dependent on dtype, tile shape, layout, and compiler decisions.

\subsection{PyTorch Baseline}

PyTorch serves as the semantic reference, correctness oracle, and practical software baseline for reporting speedup. 
Its execution path is operator dependent and may involve eager tensor operations, generic ATen kernels, compiled kernels, or optimized vendor libraries. 
Therefore, speedup over PyTorch reflects the practical benefit of a custom kernel over the available software path, while RQ3 focuses primarily on explaining the performance differences between the matched Triton and cuTile implementations.

\section{Additional Evaluation Results}
\label{appsec:additional_evaluation_results}

\subsection{RQ1: Autotuned Speedup Distributions and Category Breakdowns}
\label{appsubsec:rq1_extension}

\begin{figure*}[htbp]
    \centering
    \begin{subfigure}[b]{0.4\textwidth}
        \includegraphics[width=\textwidth]{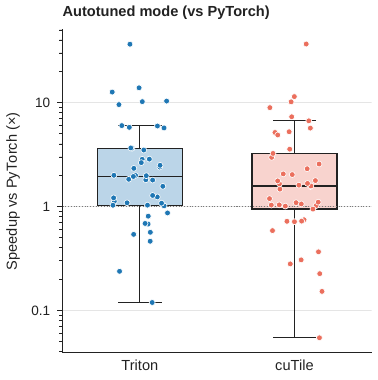}
        \caption{Autotune-mode speedup distribution across operators.}
        \label{fig:rq1ext_b}
    \end{subfigure}
    \hfill
    \begin{subfigure}[b]{0.5\textwidth}
        \includegraphics[width=\textwidth]{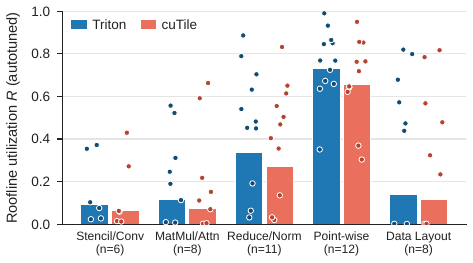}
        \caption{Per-category roofline utilization $R$ (Autotune-mode).}
        \label{fig:rq1ext_c}
    \end{subfigure}
    \caption{RQ1 extension: speedup distribution and per-category roofline utilization.}
    \label{fig:rq1ext}
\end{figure*}

\Cref{fig:rq1ext} complements the operator scatter in Section \ref{subsec:rq1}.
The median operator-level speedup is $1.95\times$ for Triton and $1.58\times$ for cuTile.
$9$ Triton implementations and $12$ cuTile implementations remain below PyTorch.
The largest operator-level speedups for both DSLs reach approximately $36\times$ on \texttt{bitonic\_sort}, whose PyTorch reference is an eager multi-pass bitonic network rather than using \texttt{torch.sort}.
The large speedup therefore reflects, in part, the overhead of the PyTorch reference execution path rather than only the efficiency of the DSL backends.
This example illustrates why the speedup distribution should be interpreted
together with the corresponding PyTorch baseline implementation.

\begin{table*}[t]
\centering
\small
\begin{tabular}{lrrrrrr}
\toprule
Category & \multicolumn{2}{c}{Speedup over PyTorch} & \multicolumn{2}{c}{Roofline Utilization} & Triton wins & cuTile wins \\
 & Triton & cuTile & Triton & cuTile & & \\
\midrule
Point-wise & $2.07\times$ & $1.84\times$ & 0.733 & 0.657 & 11 & 1 \\
Reduction/Norm & $2.56\times$ & $2.05\times$ & 0.337 & 0.269 & 9 & 2 \\
Data Layout & $3.37\times$ & $2.79\times$ & 0.139 & 0.115 & 6 & 2 \\
Stencil/Conv & $1.22\times$ & $0.86\times$ & 0.093 & 0.066 & 5 & 1 \\
Matrix Mult./Attn & $1.22\times$ & $0.79\times$ & 0.118 & 0.076 & 6 & 2 \\
\bottomrule
\end{tabular}
\caption{Category-level geometric means under autotuned mode. The last two columns count direct Triton and cuTile wins inside each category.}
\label{tab:app-rq1-category}
\end{table*}

\Cref{tab:app-rq1-category} shows that the software and hardware utilization views are complementary.
Point-wise operators obtain the highest category-level roofline utilization and exhibit many near-parity Triton and cuTile results.
Reduction / Normalization kernels also achieve strong speedups over PyTorch, often because the DSL kernels fuse operation or specialize a generation reduction template.
Data Layout operators show the largest speedups over PyTorch but much lower roofline utilization, because the category contains multi-pass sorting and selection workloads whose PyTorch references and DSL algorithms have different launch and traffic structures.
Stencil / Convolution and Matrix Multiplication / Attention remain the least utilized categories.
Their arithmetic work is substantial, but performance is frequently limited by vendor-library gaps, dynamic indexing, state placement, or pipeline formation rather than peak Tensor Core throughput alone.

Triton has the highest category-level speedup in all five groups, but this does not imply that cuTile is uniformly weaker inside a category.
cuTile wins the most informative regular tiled cases, including \texttt{matmul\_fp32\_fp16\_fp8}, \texttt{flash\_attention}, and \texttt{jacobi\_stencil\_2d}.
Its category aggregates are instead reduced by a smaller number of large deficits on irregular indexing, small reductions, sparse traversal, and runtime-controlled loops.

\subsection{RQ2: Per-operator Autotuning Gains and Roofline Headroom}
\label{appsubsec:rq2_extension}

\begin{figure}
    \centering
    \includegraphics[width=0.8\linewidth]{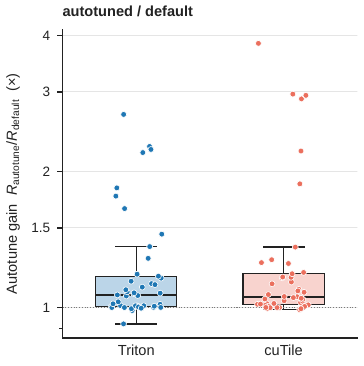}
    \caption{Triton and cuTile autotune gain $R_{\mathrm{autotune}}/R_{\mathrm{default}}$ per operator.}
    \label{fig:rq2ext}
\end{figure}

The suite-level geometric-mean tuning gain is $1.18\times$ for Triton and $1.22\times$ for cuTile, while the median gains are $1.07\times$ and $1.06\times$, respectively.
The larger cuTile geometric mean is driven by several high-gain operators rather than a uniformly larger shift.
The strongest Triton gains are \texttt{argmax} ($2.67\times$), \texttt{histogramming} ($2.27\times$), \texttt{3d\_conv} ($2.24\times$), and \texttt{2d\_conv} ($2.20\times$).
The strongest cuTile gains are \texttt{linear\_self\_attention} ($3.84\times$), \texttt{1d\_conv} ($2.97\times$), \texttt{2d\_conv} ($2.95\times$), \texttt{argmax} ($2.90\times$), and \texttt{3d\_conv} ($2.22\times$).

\begin{table}[t]
\centering
\small
\begin{tabular}{lrr}
\toprule
Metric & Triton & cuTile \\
\midrule
Default suite geometric-mean $R$ & 0.209 & 0.160 \\
Autotuned suite geometric-mean $R$ & 0.248 & 0.194 \\
Geometric-mean tuning gain & $1.18\times$ & $1.22\times$ \\
Median tuning gain & $1.07\times$ & $1.06\times$ \\
Autotuned $R\geq 0.8$ & 7/45 & 5/45 \\
Autotuned $R\geq 0.9$ & 2/45 & 1/45 \\
\bottomrule
\end{tabular}
\caption{Aggregate RQ2 results.}
\label{tab:app-rq2-summary}
\end{table}

The autotuned result is the best measured candidate in the declared search space and is not clamped to the manually selected default.
$4$ Triton operator-level results and $4$ cuTile results are therefore slightly below their defaults. 
$7$ of these $8$ regressions are within approximately $1.5\%$ of parity. 
The main exception is Triton \texttt{top\_k\_selection}, whose tuned aggregate is $0.92\times$ its default. 
We retain these values rather than replacing them with the default because they accurately describe the implemented tuning protocol.

The results separate two kinds of headroom. 
Configuration headroom is recovered by changing tile shape, warp count, pipeline stages, or occupancy. 
Backend headroom remains when the source implementation and compiler continue to generate an expensive instruction path, an unsuitable reduction organization, excessive register or shared-memory state, or an ineffective software pipeline. 
The small number of operators above $80\%$ utilization after tuning shows that the second kind of headroom dominates much of the suite.

\subsection{RQ4: Iterative Refinement Trajectories}
\label{appsubsec:rq4_extension}

The LLM track allows at most $10$ refinement iterations. A backend freezes once it is verify-clean and reaches the stopping threshold, after which it incurs no additional token cost. Failed executed iterations remain part of the token total because compilation and correctness failures are part of the search process.

\begin{table*}[!t]
\centering
\small
\begin{tabular}{lrrrr}
\toprule
Model and backend & Verify-clean coverage & Faster than PyTorch & TokenCost@10 & TokenEfficiency@10 \\
\midrule
GPT-5.5 + Triton & 45/45 & 38/45 & 0.260M & 15.83 \\
GPT-5.5 + cuTile & 43/45 & 36/43 & 0.331M & 10.79 \\
Claude Opus 4.7 + Triton & 45/45 & 38/45 & 0.277M & 12.90 \\
Claude Opus 4.7 + cuTile & 41/45 & 33/41 & 0.441M & 7.61 \\
\bottomrule
\end{tabular}
\caption{Final LLM-track coverage and geometric-mean token metrics. The faster-than-PyTorch denominator in the cuTile rows is the number of operators with at least one verify-clean implementation.}
\label{tab:app-rq4-summary}
\end{table*}

For Triton, both GPT-5.5 and Claude Opus 4.7 produce at least one verify-clean implementation for all 45 operators. 
With cuTile, GPT-5.5 achieves verify-clean coverage on 43/45 operators and Claude Opus 4.7 on 41/45; 
for the remaining 2 and 4 operators, respectively, none of the generated implementations within the 10-iteration budget passes verification.

\begin{figure}
    \centering
    \includegraphics[width=\columnwidth]{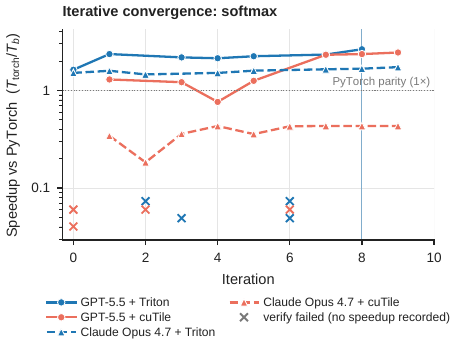}
    \caption{Iteration-level correctness-gated speedup trajectory for \texttt{softmax}. Frozen or failed backend iterations do not contribute a new valid speedup point.}
    \label{fig:rq4ext}
\end{figure}

\Cref{fig:rq4ext} illustrates why a single final score does not fully describe the refinement process. GPT-5.5 + Triton reaches a valid implementation early and peaks at approximately $2.64\times$ in iteration $8$. Claude Opus 4.7 + Triton reaches approximately $1.73\times$ by iteration $9$. GPT-5.5 + cuTile fails verification at iteration 0, later drops to approximately $0.76\times$, and reaches approximately $2.45\times$ by iteration $9$. Claude Opus 4.7 + cuTile remains near $0.43\times$. The trajectories show that later refinement can improve both backends, but the final experiment does not vary the iteration budget and therefore does not establish whether a larger budget would eliminate the backend gap.

The LLM results quantify an empirical usability difference under the TileBench protocol. They do not isolate whether the cause is API expressibility, primitive availability, compiler constraints, documentation, public examples, or prior model exposure. We therefore avoid treating model familiarity as a demonstrated causal explanation.

\section{Computational Budget}
All experiments were run on a single NVIDIA B200 GPU with 180GB HBM3e memory.
The main benchmark, autotuning, NCU report generation experiments consumed approximately 100 GPU-hours in total.

\section{LLM API Usage}
\label{llm_api_usage}
For the iterative LLM code-generation track, GPT-5.5 consumed $18.68$M API tokens with a total cost of $\$279.05$, while Claude Opus 4.7 consumed $21.09$M API tokens with a total cost of $\$191.66$. In total, the LLM track consumed $39.77$M API tokens and cost $\$470.71$.
Because both evaluated LLMs are closed-source API models, their parameter counts are not publicly available.

\section{Artifact Sources and Licenses}
\label{artifact_sources_and_licenses}

TileBench uses TritonBench and LeetGPU as sources of operator coverage and task semantics. 
TritonBench is distributed under the Apache License 2.0. For TritonBench-derived components, we preserve the original attribution and license notices and document our modifications when applicable.

LeetGPU is distributed under the Creative Commons Attribution-NonCommercial-NoDerivatives 4.0 International license (CC BY-NC-ND 4.0). This license permits non-commercial sharing with attribution, but does not permit redistribution of adapted material. Accordingly, TileBench does not redistribute LeetGPU source code, test files, or modified problem statements. For operators selected from LeetGPU challenge topics, we provide independently written PyTorch references, Triton implementations, cuTile implementations, and TileBench configuration files, and we cite the original LeetGPU artifact.

TileBench artifacts are newly written by the authors. Third-party artifacts, tools, and dependencies remain governed by their original licenses and terms of use.

\end{document}